\documentclass[%
reprint,
superscriptaddress,
 amsmath,amssymb,
 prl
]{revtex4-2}

\usepackage{xcolor}
\usepackage{textgreek}
\usepackage{graphicx}
\usepackage{dcolumn}
\usepackage{bm}

\begin{document}

\preprint{APS/PRL}

\title{Controlled Growth of the Self-Modulation of a Relativistic Proton Bunch in Plasma}

\author{L.~Verra}
\email{livio.verra@cern.ch}
\affiliation{CERN, Geneva, Switzerland}
\affiliation{Max Planck Institute for Physics, Munich, Germany}
\affiliation{Technical University Munich, Munich, Germany}
\author{G.~Zevi Della Porta}
\affiliation{CERN, Geneva, Switzerland}
\author{J.~Pucek}
\affiliation{Max Planck Institute for Physics, Munich, Germany}
\affiliation{Technical University Munich, Munich, Germany}
\author{T.~Nechaeva}
\affiliation{Max Planck Institute for Physics, Munich, Germany}
\affiliation{Technical University Munich, Munich, Germany}
\author{S.~Wyler}
\affiliation{Ecole Polytechnique Federale de Lausanne (EPFL), Swiss Plasma Center (SPC), Lausanne, Switzerland}
\author{M.~Bergamaschi} 
\affiliation{Max Planck Institute for Physics, Munich, Germany}
\author{E.~Senes}
\affiliation{CERN, Geneva, Switzerland}
\author{E.~Guran}
\affiliation{CERN, Geneva, Switzerland}
\author{J.T.~Moody}
\affiliation{Max Planck Institute for Physics, Munich, Germany}
\author{M.Á.~Kedves}
\affiliation{Wigner Research Centre for Physics, Budapest, Hungary}
\author{E.~Gschwendtner}
\affiliation{CERN, Geneva, Switzerland}
\author{P.~Muggli}
\affiliation{Max Planck Institute for Physics, Munich, Germany}
\collaboration{AWAKE Collaboration}
\noaffiliation
\author{R.~Agnello}
\affiliation{Ecole Polytechnique Federale de Lausanne (EPFL), Swiss Plasma Center (SPC), Lausanne, Switzerland}
\author{C.C.~Ahdida}
\affiliation{CERN, Geneva, Switzerland}
\author{M.C.A.~Goncalves}
\affiliation{CERN, Geneva, Switzerland}
\author{Y.~Andrebe}
\affiliation{Ecole Polytechnique Federale de Lausanne (EPFL), Swiss Plasma Center (SPC), Lausanne, Switzerland}
\author{O.~Apsimon}
\affiliation{University of Liverpool, Liverpool, UK}
\affiliation{Cockcroft Institute, Daresbury, UK}
\author{R.~Apsimon}
\affiliation{Cockcroft Institute, Daresbury, UK} 
\affiliation{Lancaster University, Lancaster, UK}
\author{J.M.~Arnesano}
\affiliation{CERN, Geneva, Switzerland}
\author{A.-M.~Bachmann}
\affiliation{Max Planck Institute for Physics, Munich, Germany}
\author{D.~Barrientos}
\affiliation{CERN, Geneva, Switzerland}
\author{F.~Batsch}
\affiliation{Max Planck Institute for Physics, Munich, Germany}
\author{V.~Bencini}
\affiliation{CERN, Geneva, Switzerland}
\affiliation{John Adams Institute, Oxford University, Oxford, UK}
\author{P.~Blanchard}
\affiliation{Ecole Polytechnique Federale de Lausanne (EPFL), Swiss Plasma Center (SPC), Lausanne, Switzerland}
\author{P.N.~Burrows}
\affiliation{John Adams Institute, Oxford University, Oxford, UK}
\author{B.~Buttensch{\"o}n}
\affiliation{Max Planck Institute for Plasma Physics, Greifswald, Germany}
\author{A.~Caldwell}
\affiliation{Max Planck Institute for Physics, Munich, Germany}
\author{J.~Chappell}
\affiliation{UCL, London, UK}
\author{E.~Chevallay}
\affiliation{CERN, Geneva, Switzerland}
\author{M.~Chung}
\affiliation{UNIST, Ulsan, Republic of Korea}
\author{D.A.~Cooke}
\affiliation{UCL, London, UK}
\author{C.~Davut}
\affiliation{Cockcroft Institute, Daresbury, UK} 
\affiliation{University of Manchester, Manchester, UK}
\author{G.~Demeter}
\affiliation{Wigner Research Centre for Physics, Budapest, Hungary}
\author{A.C.~Dexter}
\affiliation{Cockcroft Institute, Daresbury, UK} 
\affiliation{Lancaster University, Lancaster, UK}
\author{S.~Doebert}
\affiliation{CERN, Geneva, Switzerland}
\author{F.A.~Elverson}
\affiliation{CERN, Geneva, Switzerland}
\author{J.~Farmer}
\affiliation{CERN, Geneva, Switzerland}
\affiliation{Max Planck Institute for Physics, Munich, Germany}
\author{A.~Fasoli}
\affiliation{Ecole Polytechnique Federale de Lausanne (EPFL), Swiss Plasma Center (SPC), Lausanne, Switzerland}
\author{V.~Fedosseev}
\affiliation{CERN, Geneva, Switzerland}
\author{R.~Fonseca}
\affiliation{ISCTE - Instituto Universit\'{e}ario de Lisboa, Portugal} 
\affiliation{GoLP/Instituto de Plasmas e Fus\~{a}o Nuclear, Instituto Superior T\'{e}cnico, Universidade de Lisboa, Lisbon, Portugal}
\author{I.~Furno}
\affiliation{Ecole Polytechnique Federale de Lausanne (EPFL), Swiss Plasma Center (SPC), Lausanne, Switzerland}
\author{A.~Gorn}
\affiliation{Budker Institute of Nuclear Physics SB RAS, Novosibirsk, Russia}
\affiliation{Novosibirsk State University, Novosibirsk, Russia}
\author{E.~Granados}
\affiliation{CERN, Geneva, Switzerland}
\author{M.~Granetzny}
\affiliation{University of Wisconsin, Madison, Wisconsin, USA}
\author{T.~Graubner}
\affiliation{Philipps-Universit{\"a}t Marburg, Marburg, Germany}
\author{O.~Grulke}
\affiliation{Max Planck Institute for Plasma Physics, Greifswald, Germany}
\affiliation{Technical University of Denmark, Lyngby, Denmark}
\author{V.~Hafych}
\affiliation{Max Planck Institute for Physics, Munich, Germany}
\author{J.~Henderson}
\affiliation{Cockcroft Institute, Daresbury, UK}
\affiliation{Accelerator Science and Technology Centre, ASTeC, STFC Daresbury Laboratory, Warrington, UK}
\author{M.~H{\"u}ther}
\affiliation{Max Planck Institute for Physics, Munich, Germany}
\author{V.~Khudiakov}
\affiliation{Heinrich-Heine-Universit{\"a}t D{\"u}sseldorf, D{\"u}sseldorf, Germany}
\affiliation{Budker Institute of Nuclear Physics SB RAS, Novosibirsk, Russia}
\author{S.-Y.~Kim}
\affiliation{UNIST, Ulsan, Republic of Korea}
\affiliation{CERN, Geneva, Switzerland}
\author{F.~Kraus}
\affiliation{Philipps-Universit{\"a}t Marburg, Marburg, Germany}
\author{M.~Krupa}
\affiliation{CERN, Geneva, Switzerland}
\author{T.~Lefevre}
\affiliation{CERN, Geneva, Switzerland}
\author{L.~Liang}
\affiliation{Cockcroft Institute, Daresbury, UK}
\affiliation{University of Manchester, Manchester, UK}
\author{S.~Liu}
\affiliation{TRIUMF, Vancouver, Canada}
\author{N.~Lopes}
\affiliation{GoLP/Instituto de Plasmas e Fus\~{a}o Nuclear, Instituto Superior T\'{e}cnico, Universidade de Lisboa, Lisbon, Portugal}
\author{K.~Lotov}
\affiliation{Budker Institute of Nuclear Physics SB RAS, Novosibirsk, Russia}
\affiliation{Novosibirsk State University, Novosibirsk, Russia}
\author{M.~Martinez~Calderon}
\affiliation{CERN, Geneva, Switzerland}
\author{S.~Mazzoni}
\affiliation{CERN, Geneva, Switzerland}
\author{D.~Medina~Godoy} 
\affiliation{CERN, Geneva, Switzerland}
\author{K.~Moon}
\affiliation{UNIST, Ulsan, Republic of Korea}
\author{P.I.~Morales~Guzm\'{a}n}
\affiliation{Max Planck Institute for Physics, Munich, Germany}
\author{M.~Moreira}
\affiliation{GoLP/Instituto de Plasmas e Fus\~{a}o Nuclear, Instituto Superior T\'{e}cnico, Universidade de Lisboa, Lisbon, Portugal}
\author{E.~Nowak}
\affiliation{CERN, Geneva, Switzerland}
\author{C.~Pakuza}
\affiliation{John Adams Institute, Oxford University, Oxford, UK}
\author{H.~Panuganti}
\affiliation{CERN, Geneva, Switzerland}
\author{A.~Pardons}
\affiliation{CERN, Geneva, Switzerland}
\author{K.~Pepitone}
\affiliation{Angstrom Laboratory, Department of Physics and Astronomy, Uppsala, Sweden}
\author{A.~Perera}
\affiliation{Cockcroft Institute, Daresbury, UK}
\affiliation{University of Liverpool, Liverpool, UK}
\author{A.~Pukhov}
\affiliation{Heinrich-Heine-Universit{\"a}t D{\"u}sseldorf, D{\"u}sseldorf, Germany}
\author{R.L.~Ramjiawan}
\affiliation{CERN, Geneva, Switzerland}
\affiliation{John Adams Institute, Oxford University, Oxford, UK}
\author{S.~Rey}
\affiliation{CERN, Geneva, Switzerland}
\author{O.~Schmitz}
\affiliation{University of Wisconsin, Madison, Wisconsin, USA}
\author{F.~Silva}
\affiliation{INESC-ID, Instituto Superior Técnico, Universidade de Lisboa, Lisbon, Portugal}
\author{L.~Silva}
\affiliation{GoLP/Instituto de Plasmas e Fus\~{a}o Nuclear, Instituto Superior T\'{e}cnico, Universidade de Lisboa, Lisbon, Portugal}
\author{C.~Stollberg}
\affiliation{Ecole Polytechnique Federale de Lausanne (EPFL), Swiss Plasma Center (SPC), Lausanne, Switzerland}
\author{A.~Sublet}
\affiliation{CERN, Geneva, Switzerland}
\author{C.~Swain}
\affiliation{Cockcroft Institute, Daresbury, UK}
\affiliation{University of Liverpool, Liverpool, UK}
\author{A.~Topaloudis}
\affiliation{CERN, Geneva, Switzerland}
\author{N.~Torrado}
\affiliation{GoLP/Instituto de Plasmas e Fus\~{a}o Nuclear, Instituto Superior T\'{e}cnico, Universidade de Lisboa, Lisbon, Portugal}
\author{P.~Tuev}
\affiliation{Budker Institute of Nuclear Physics SB RAS, Novosibirsk, Russia}
\affiliation{Novosibirsk State University, Novosibirsk, Russia}
\author{F.~Velotti}
\affiliation{CERN, Geneva, Switzerland}
\author{V.~Verzilov}
\affiliation{TRIUMF, Vancouver, Canada}
\author{J.~Vieira}
\affiliation{GoLP/Instituto de Plasmas e Fus\~{a}o Nuclear, Instituto Superior T\'{e}cnico, Universidade de Lisboa, Lisbon, Portugal}
\author{M.~Weidl}
\affiliation{Max Planck Institute for Plasma Physics, Munich, Germany}
\author{C.~Welsch}
\affiliation{Cockcroft Institute, Daresbury, UK}
\affiliation{University of Liverpool, Liverpool, UK}
\author{M.~Wendt}
\affiliation{CERN, Geneva, Switzerland}
\author{M.~Wing}
\affiliation{UCL, London, UK}
\author{J.~Wolfenden}
\affiliation{Cockcroft Institute, Daresbury, UK}
\affiliation{University of Liverpool, Liverpool, UK}
\author{B.~Woolley}
\affiliation{CERN, Geneva, Switzerland}
\author{G.~Xia}
\affiliation{Cockcroft Institute, Daresbury, UK}
\affiliation{University of Manchester, Manchester, UK}
\author{V.~Yarygova}
\affiliation{Budker Institute of Nuclear Physics SB RAS, Novosibirsk, Russia}
\affiliation{Novosibirsk State University, Novosibirsk, Russia}
\author{M.~Zepp}
\affiliation{University of Wisconsin, Madison, Wisconsin, USA}

\date{\today}

\begin{abstract}

A long, narrow, relativistic charged particle bunch propagating in plasma is subject to the self-modulation (SM) instability. %
We show that SM of a proton bunch can be seeded by the wakefields driven by a preceding electron bunch. %
SM timing reproducibility and control are at the level of a small fraction of the modulation period.
With this seeding method, we independently control the amplitude of the seed wakefields with the charge of the electron bunch and the growth rate of SM with the charge of the proton bunch. 
Seeding leads to larger growth of the wakefields than in the instability case.

\end{abstract}

\maketitle

\par \textit{Introduction.---} Instabilities are of paramount importance in plasma physics~\cite{instabilities}. %
Similar instabilities occur in vastly different plasmas, from astrophysical~\cite{astro,astroneutrino}, to laboratory~\cite{lab} and fusion~\cite{inertialconffus}, to quantum~\cite{quantum} and even to quark-gluon plasmas~\cite{quarkgluon}. %
They can be disruptive and must then be suppressed, or beneficial and must then be controlled. %
Charged particle beams propagating in plasma are subject to a number of instabilities, including different occurrences of the two-stream instability~\cite{Farley,buneman}. %
In the case of a long, narrow, relativistic charged particle bunch, the instability is transverse and it is called the self-modulation instability (SMI)~\cite{KUMAR:GROWTH}. %

\par Relativistic charged particle bunches traveling in plasma leave behind a perturbation in the plasma electron density. 
This perturbation provides a restoring force that induces an oscillation of the plasma electrons with angular frequency $\omega_{pe}=\sqrt{\frac{n_{pe} e^2}{m_e \varepsilon_0}}$, where $n_{pe}$ is the plasma electron density, $e$ and $m_e$ are the electron charge and mass, $\varepsilon_0$ is the vacuum permittivity. %
The local charge non-neutrality sustains fields with transverse and longitudinal components, known as wakefields, that can have amplitudes appealing for high-gradient particle acceleration~\cite{TAJIMA:LWFA,PWFA:CHEN}. %

\par SMI~\cite{KUMAR:GROWTH} develops when the bunch duration %
is much longer than the period of the wakefields: ${\sigma_t\gg T_{pe}=2\pi/\omega_{pe}}$. %
Transverse wakefields act back on the bunch, modulating its radius and thus its charge density. %
The modulated distribution drives enhanced wakefields, causing the growth of SMI that, at saturation, leaves the long bunch fully modulated into a train of microbunches 
with periodicity $\sim T_{pe}$.
The timing of the microbunches along the train is tied to that of the wakefields since microbunches develop in their focusing phase. %

\par When a long proton ($p^+$) bunch enters a pre-ionized plasma, SMI develops from the wakefields driven by noise~\cite{KOSTANTIN:NOISE} or by imperfections in the incoming bunch charge distribution~\cite{FABIAN:PRL}. %
Thus, the initial conditions vary from event to event and so do the timing and amplitude of the wakefields. %
However, the outcome can be controlled by seeding the instability, i.e., by fixing the initial conditions from which the instability grows. %

\par Seeding requires driving initial transverse wakefields with amplitude larger than those driven by the noise or imperfections in the bunch so that the self-modulation (SM) develops from a well-defined time, and with well-defined initial amplitude and growth rate. %
We demonstrated experimentally that a high-energy, long $p^+$ bunch undergoes SMI when traveling in plasma~\cite{KARL:PRL}, and that the resulting microbunch train resonantly excites large amplitude wakefields~\cite{MARLENE:PRL,AW:NATURE}.
A relativistic ionization front (RIF) generating the plasma and co-propagating within the $p^+$ bunch can provide the seed by the rapid onset of the beam-plasma interaction~\cite{FABIAN:PRL}. %
In this case, the amplitude of the seed wakefields and the growth rate of SM depend on the $p^+$ bunch density at the RIF and cannot be varied independently. %
Moreover, the front of the bunch propagates as if in vacuum and thus remains unmodulated.

\par The initial transverse seed wakefields can also be provided by a preceding charged particle bunch~\cite{MATHIAS:ESSM, PATRIC:ESSM}. %
In this case, seeding amplitude and growth rate of SM can be varied independently. %
Moreover, as the seed wakefields act on the whole $p^+$ bunch, the entire bunch self-modulates. %

\par The protons that are defocused out of the wakefields are probes for the amplitude of the wakefields at early distances along plasma, during SM growth, before saturation~\cite{MARLENE:PRL,MARLENE:PRAB}. %
Theoretical and numerical simulation results~\cite{KUMAR:GROWTH,PUKHOV:GROWTH,doi:10.1063/1.4933129,SCH:GROWTH}
show that, in the linear regime, the amplitude of the transverse wakefields along the bunch ($t$) and along the plasma ($z$) grows as ${W_{\perp }(t,z)= W_{\perp 0}\exp\left(\Gamma(t,z)z\right)}$. %
In the case of seeding with an electron ($e^-$) bunch, the amplitude of the initial wakefields $W_{\perp 0}(z=0)$ depends solely on the $e^-$ bunch parameters, while the growth rate of SM $\Gamma(t,z)$ depends solely on those of the $p^+$ bunch.
The radial extent reached by defocused protons a distance downstream of the plasma is proportional to the transverse momentum they acquire from these wakefields, and therefore increases with the growth of SM.

\par In this \textit{Letter}, we demonstrate with experimental results that SM of a long, relativistic $p^+$ bunch in plasma can be seeded by a preceding $e^-$ bunch.
We show that the growth of SM increases when increasing the charge of the seed $e^-$ bunch $Q_e$ or the charge of the $p^+$ bunch $Q_p$.
We attribute these changes to a change in the amplitude of the transverse seed wakefields $W_{\perp0}(Q_e)$ or in the SM growth rate $\Gamma(Q_p)$. %
These observations are possible because the $e^-$ bunch effectively seeds SM and they are in agreement with theoretical and simulation predictions~\cite{KUMAR:GROWTH, PUKHOV:GROWTH, SCH:GROWTH,Lotov_2021}. 
When seeding, the growth of the process is larger  than in the SMI case~\cite{KUMAR:GROWTH}.
We also observe adiabatic focusing of the front of the $p^+$ bunch, where the growth of SM is small. %
In addition, $e^-$ bunch seeding allows for the timing of the process to be controlled at the sub-modulation-period, picosecond time scale. %
\begin{figure}[h!]
\centering
\includegraphics[scale=0.4]{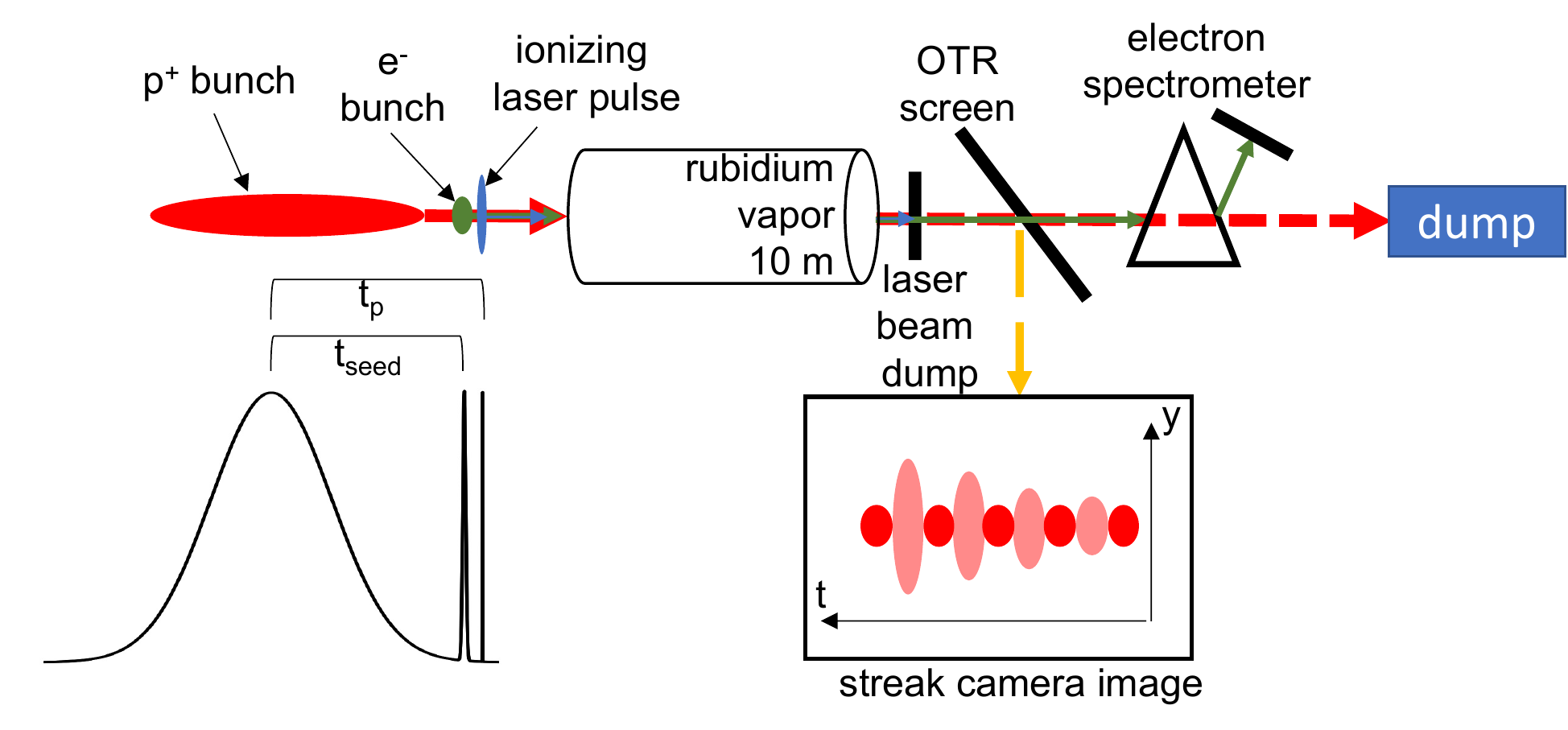}
\caption{Schematic of the experimental setup: the ionizing laser pulse enters the vapor source $t_p$ ahead of the $p^+$ bunch center and ionizes the rubidium atoms, creating the plasma. %
The seed $e^-$ bunch follows, $t_{\mathrm{seed}}$ ahead of the $p^+$ bunch. 
The optical transition radiation produced at a screen positioned $3.5\,$m downstream of the plasma exit is imaged on the entrance slit of a streak camera.
A schematic example of a time-resolved image of the self-modulated $p^+$ bunch provided by the streak camera is shown in the inset. %
The magnetic spectrometer is located downstream of the screen. %
}
\label{fig:1}
\end{figure}
\par \textit{Experimental Setup.---} The measurements took place in the context of the AWAKE experiment~\cite{PATRIC:READINESS}, whose goal is to accelerate $e^-$ bunches to GeV energies, ultimately for high-energy physics applications~\cite{,ALLEN:EPCOLL}.  

Figure~\ref{fig:1} shows a schematic of the experimental setup. %
A $10$-m-long source provides rubidium vapor density adjustable in the $n_{\mathrm{vap}} = (0.5-10) \cdot 10^{14}\,$cm$\textsuperscript{-3}$ range~\cite{PATRIC:READINESS}. %
The density is measured to better than $0.5\%$~\cite{FABIAN:DENSITY} at the source ends. %
An $\sim 120\,$fs, $\sim 100\,$mJ laser pulse ($\lambda=780\,$nm) produces a RIF that creates the plasma by ionizing the vapor (RbI$\rightarrow$RbII). %
Previous experiments~\cite{KARL:PRL} showed that the RIF ionizes $\sim100\%$ of the vapor along its path, producing an $\sim 2$-mm-diameter plasma column with density equal to that of the vapor. %
The RIF is placed $t_p=620\,$ps$\ (\sim 2.5\, \sigma_t$) ahead of the center of the $400\,$GeV/c, $\sigma_t\sim240\,$ps,  $p^+$ bunch provided by the CERN SPS. %
Therefore, it does not seed SM~\cite{FABIAN:PRL}. %
The $p^+$ bunch is synchronized with the RIF with root mean square (rms) variation of $15\,$ps$\ll\sigma_t$, which is therefore negligible. %

\par Optical transition radiation (OTR) is emitted when protons enter an aluminum-coated silicon wafer, positioned $3.5\,$m downstream of the plasma exit.
The OTR is imaged onto the entrance slit of a streak camera that provides time-resolved images of the charge density distribution of the $p^+$ bunch ($t,y$)~\cite{KARL:STREAK} in a $\sim 180$-\textmu m-wide slice (the spatial resolution of the optical system) near the propagation axis.
The streak camera temporal resolution is $\sim 2\,$ps in the $210\,$ps time window, sufficient to resolve the microbunch train as the plasma period is $T_{pe}=11.04$ and $11.38\,$ps, for the values of $n_{pe}$ used in this experiment. %
An ultraviolet pulse derived from the same laser oscillator as that producing the RIF generates an $18.3\,$MeV $e^-$ bunch in a photo-injector and booster cavity~\cite{PEPITONE201673}. 
The $e^-$ bunch and the RIF have a relative rms timing jitter $<1\,$ps ($\ll T_{pe}$)~\cite{ANNA:THESIS}. %
The delay $t_{\mathrm{seed}}$ between the $e^-$ and the $p^+$ bunch centers can be adjusted using a translation stage. %
We use a magnetic spectrometer~\cite{BAUCHE2019103} to measure the energy spectrum of the $e^-$ bunch after propagation with and without plasma~\cite{LIVIO:EPS}. %

\par We use a bleed-through of the ionizing laser pulse, thus also synchronized with the $e^-$ bunch at the sub-ps time scale~\cite{FABIAN:MARKER}, to determine on the time-resolved images the bunch train timing with respect to that of the $e^-$ bunch. 
This is necessary to circumvent the $\sim 5\,$ps rms jitter (${\sim T_{pe}/2}$) of the triggering system.

\par
\begin{figure}[h!]
\centering
\includegraphics[scale=0.25]{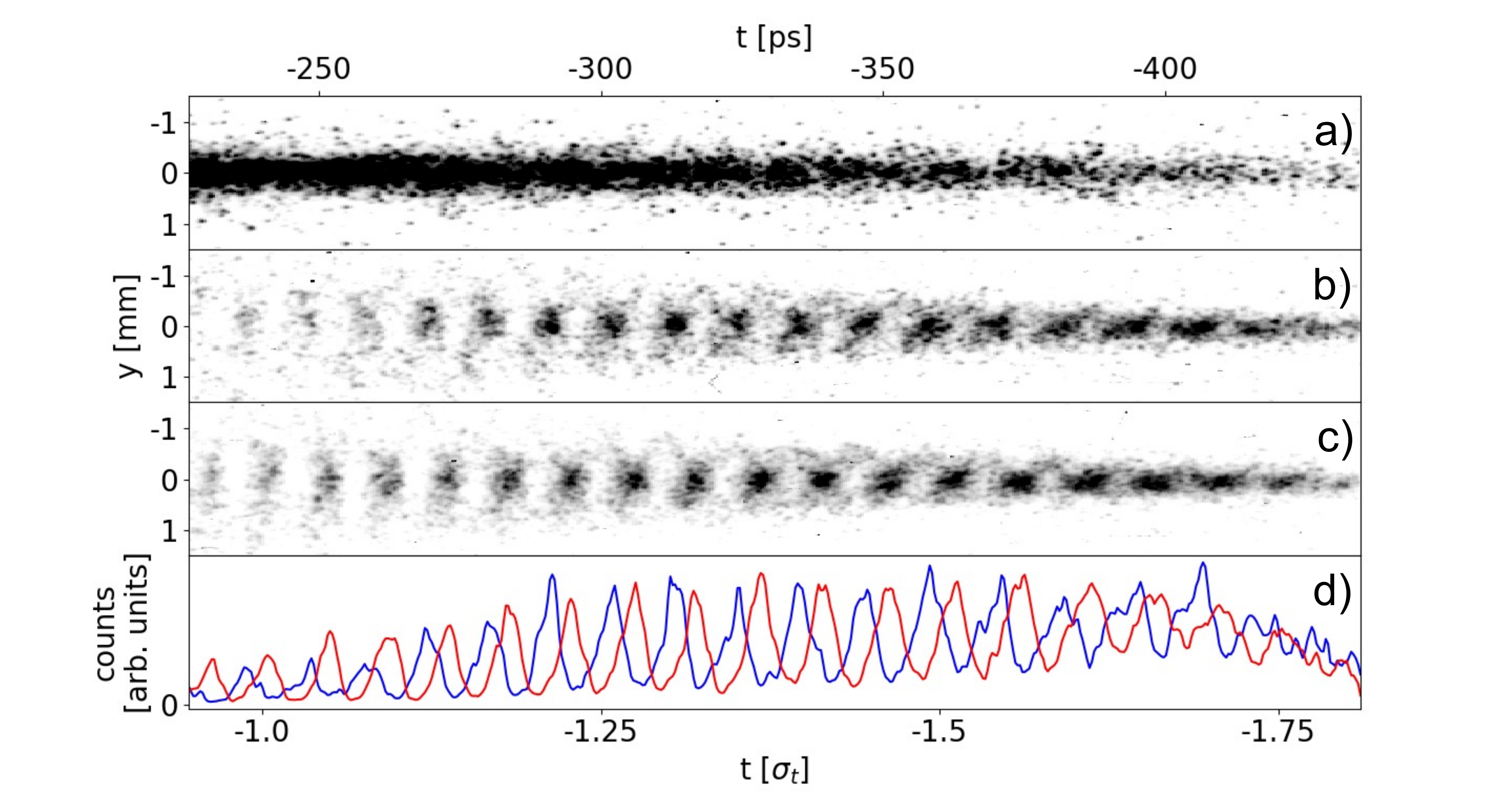}
\caption{
Time-resolved images ($t,y$) of the $p^+$ bunch at the OTR screen obtained by averaging ten single-event images ($210\,$ps, $Q_p=14.7\,$nC).
Bunch center at $t=0\,$ps, the bunch travels from left to right.
Horizontal axis: time along the bunch normalized to the incoming bunch duration $\sigma_t$. %
a) No plasma (incoming bunch).
b) Plasma ($n_{pe}=1.02\cdot10^{14}\,$cm$^{-3}$) and $e^-$ bunch with $Q_e=249\,$pC, $t_{\mathrm{seed}}=614\,$ps ahead of the $p^+$ bunch center.
c) Same as (b) but $e^-$ bunch delayed by $6.7\,$ps ($t_{\mathrm{seed}}=607.3\,$ps).
All images have the same color scale.
d) On-axis time profiles of (b) (blue line) and (c) (red line) obtained by summing counts over $-0.217\leq y \leq0.217\,$mm.
}
\label{fig:2}
\end{figure}

 \textit{Experimental Results.---} We first present a new and important result that is necessary for the measurements presented hereafter: the seeding of SM by the $e^-$ bunch. %
The incoming $p^+$ bunch with $Q_p=(14.7\pm0.2)\,$nC has a continuous charge distribution (Fig.~\ref{fig:2}(a), no plasma) with an approximately 2D-Gaussian ($t,y$) charge density profile.
With the plasma ($n_{pe}=1.02\cdot10^{14}\,$cm$^{-3}$) and the $Q_e=(249\pm17)\,$pC $e^-$ bunch placed $t_{\mathrm{seed}}=612\,$ps ahead of the center of the $p^+$ bunch, we observe the clear formation of a train of microbunches on the image resulting from the average of ten consecutive single-event images (Fig.~\ref{fig:2}(b)). %
This indicates that SM is reproducible from event to event. %
The period of the modulation is $11.3\,$ps, close to $T_{pe}$ as expected from SM~\cite{KUMAR:GROWTH,KARL:PRL}. %
We measure the timing variation of the microbunch train with respect to the $e^-$ bunch by performing a discrete Fourier transform (DFT, see Supplemental Material of~\cite{FABIAN:PRL}) analysis of the on-axis %
time profile of single-event images. %
The rms timing variation is $\Delta t_{rms}= 0.06\,T_{pe}$, demonstrating that the $e^-$ bunch effectively seeds SM. %
The same measurement without the $e^-$ bunch yields $\Delta t_{rms}= 0.26\,T_{pe}$, consistent with uniform variation of the timing over $T_{pe}$ ($\Delta t_{rms}= 0.29\,T_{pe}$), confirming the occurrence of SMI, as was also observed in~\cite{FABIAN:PRL}. %
\par We also observe seeding of SM with ${Q_p=(46.9\pm0.5)\,}$nC and the same value of $Q_e=249\,$pC, i.e., with $p^+$ bunch and plasma parameters similar to those of~\cite{FABIAN:PRL}. %
This indicates that the $e^-$ bunch drives transverse wakefields with amplitude exceeding the seeding threshold value of $(2.8-4.0)\,$MV/m, determined in~\cite{FABIAN:PRL} when seeding with RIF. %
The amplitude thus also exceeds that for the lower $Q_p=14.7\,$nC (Fig.~\ref{fig:2}) since the %
seeding threshold is expected to scale with $Q_p$. %

\par Figure~\ref{fig:2}(c) shows an averaged time-resolved image obtained after delaying the seed $e^-$ bunch timing by $6.7\,$ps with respect to the case of Fig.~\ref{fig:2}(b). %
The bunch train is again clearly visible and timing analysis shows an rms variation of $0.07\,T_{pe}$, confirming the seeding of SM. %
Figure~\ref{fig:2}(d) shows that the temporal profiles of Fig.~\ref{fig:2}(c) (red curve) is shifted in time by $(7.2\pm1.0)\,$ps with respect to that of Fig.~\ref{fig:2}(b) (blue curve). %
This demonstrates that the timing of the $p^+$ bunch modulation and thus also the timing of the wakefields are tied to that of the seed within a small fraction of $T_{pe}$. %

\par
\begin{figure}[h!]
\centering
\includegraphics[scale=0.25]{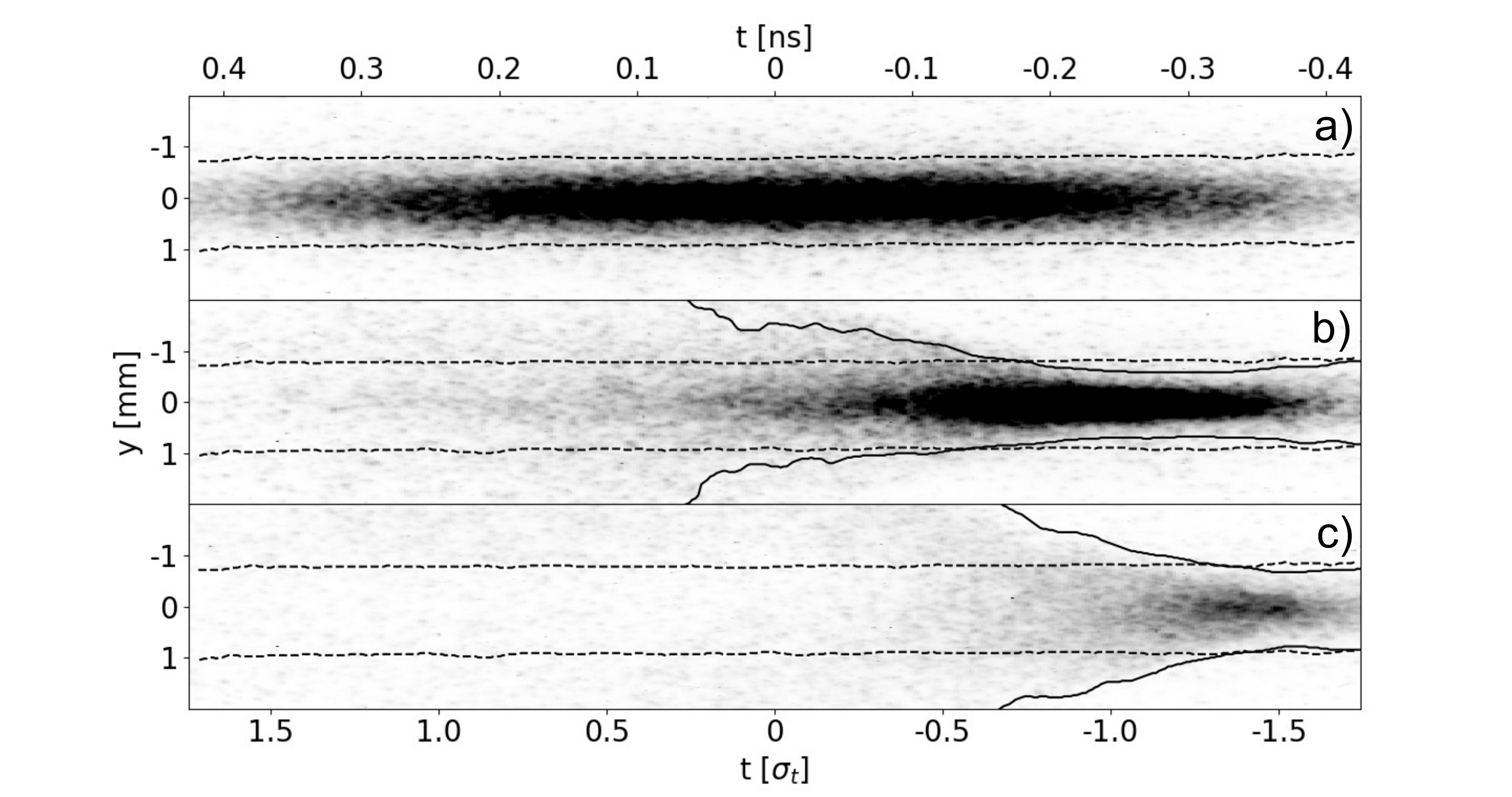}
\caption{
Time-resolved images ($t,y$) of the $p^+$ bunch ($1.1\,$ns, $Q_p=14.7\,$nC) obtained by averaging ten single-event images. %
a) No plasma (incoming bunch). %
b) Plasma (${n_{pe}=0.97\cdot10^{14}\,}$cm$^{-3}$) and no $e^-$ bunch (SMI). %
c) Plasma and $e^-$ bunch with $Q_e=249\,$pC (seeded SM). %
All images have the same color scale. %
Black dashed lines in (a) and continuous lines in (b) and (c) indicate, for each time-column of the images, the points where the transverse distribution reaches $20\%$ of its peak value.
The distance between the lines is the transverse extent $w_{\mathrm{off}}$ (a) and $w$ (b, c). %
Dashed lines of (a) also plotted in (b) and (c) for reference. %
}
\label{fig:3}
\end{figure}

 As the amplitude of the wakefields grows along the bunch and along the plasma~\cite{KARL:PRL,MARLENE:PRL}, one may expect them to produce a smaller size of the successive microbunches at the plasma exit, possibly also with larger emittance due to the nonlinear nature of the transverse wakefields. %
These %
two effects are the likely causes for the increase in transverse size of the microbunches along the train observed in Figs.~\ref{fig:2}(b) and (c), as the OTR screen is positioned $3.5\,$ m downstream of the plasma exit. %
We also note that the $p^+$ bunch self-modulates starting from the visible front of the bunch ($t>-1.82\,\sigma_t$ on Figs.~\ref{fig:2}(b) and (c)), as the seed wakefields act on the entire bunch, and that on these figures the charge density at the bunch front is higher than in the case without plasma (Fig.~\ref{fig:2}(a)).
This is due to the focusing associated with the formation of the microbunches and to global plasma adiabatic focusing (see Fig.~\ref{fig:3}).

\par We measure the transverse extent of the $p^+$ bunch distribution along the bunch on $1.1\,$ns, time-resolved images (Fig.~\ref{fig:3}). %
We define this extent $w$ for each time-column of the image as the distance between the two points ($\pm y$) where the transverse distribution reaches $20\%$ of its peak value, when detectable. %
In the case without plasma (Fig.~\ref{fig:3}(a), incoming bunch), $w_{\mathrm{off}}=1.7\,$mm is constant along the bunch (black dashed line) and corresponds to the $\sigma_y\sim 0.37\,$mm rms size of the bunch at the screen.
\par In the case with plasma (hereafter $n_{pe}=0.97\cdot10^{14}\,$cm$^{-3}$) and no $e^-$ bunch (SMI~\cite{FABIAN:PRL}, Fig.~\ref{fig:3}(b)), the transverse extent (black continuous lines) indicates that first the effect of plasma adiabatic focusing dominates, i.e., $w$ decreases, due to the cancellation of the $p^+$ bunch space-charge field by the plasma electrons (${t<-0.8\,\sigma_t}$). %
Then the effect of defocusing due to SM development dominates and $w$ increases.

We note that the time resolution of these images is not sufficient to evidence the microbunch structure and the charge distribution appears continuous along the bunch.

\par In the case with plasma and $e^-$ bunch ($Q_e=249\,$pC, seeded SM, Fig.~\ref{fig:3}(c), all other parameters kept constant), the same focusing effect as in the SMI case first dominates, but the %
effect of defocusing starts earlier: $t\sim -1.5\,\sigma_t$ rather than $t\sim -0.8\,\sigma_t$ (Fig.~\ref{fig:3}(b)).

\par
\begin{figure}[h!]
\centering
\includegraphics[scale=0.25]{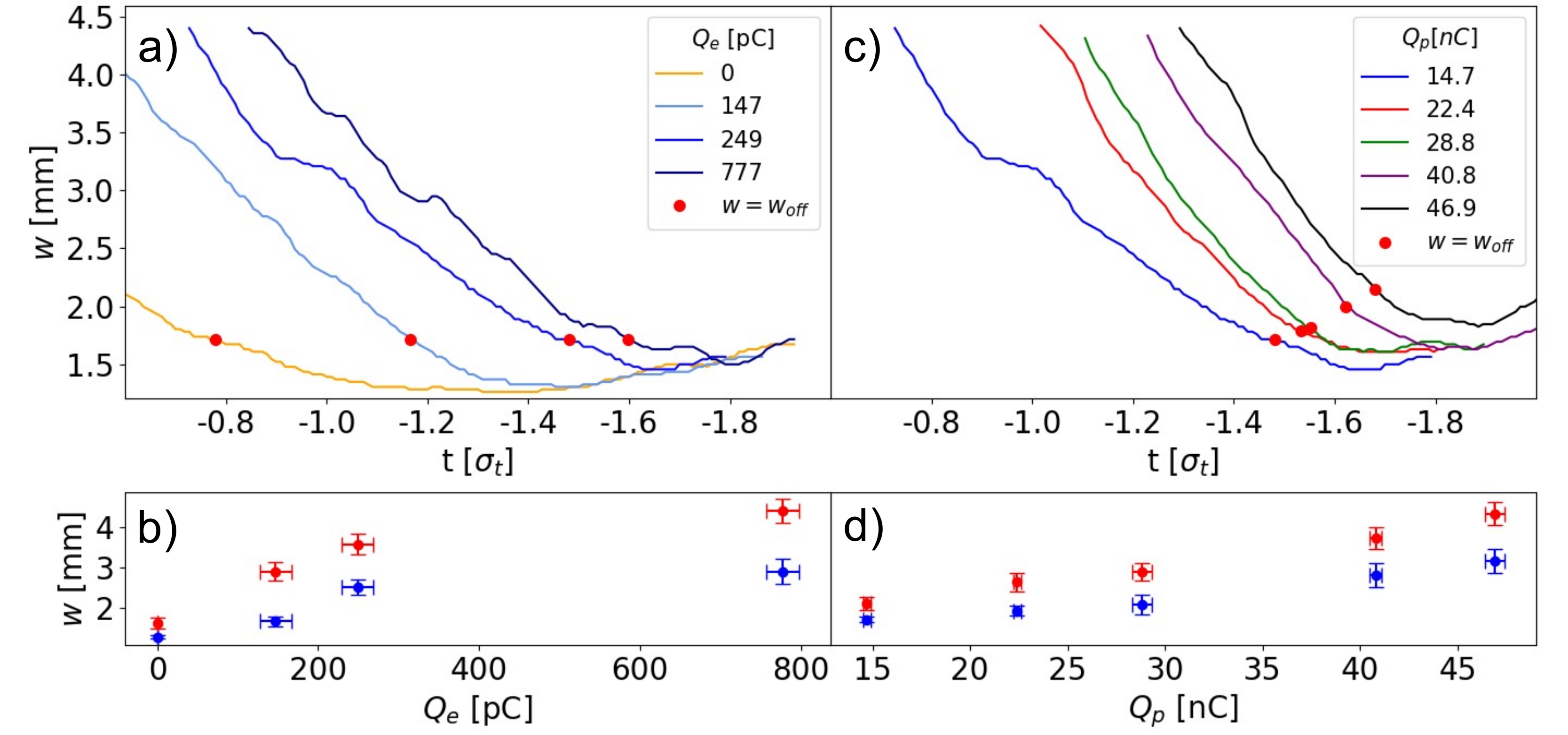}
\caption{Top row: transverse extent $w$ along the $p^+$ bunch as a function of time along the bunch normalized to the incoming bunch duration $\sigma_t$. %
a) Varying the $e^-$ bunch charge (see legend), $Q_e=0$ (SMI), $Q_e>0$ (seeded SM), $Q_p=14.7\,$nC. %
c) Varying the $p^+$ bunch charge $Q_p$ (see legend), $Q_e=249\,$pC.
Red points indicate the time along the bunch when $w=w_{\mathrm{off}}$.
Bottom row: b) $w$ as a function of $Q_e$ at $t=-1.19$ (blue points) and $t=-0.84\,\sigma_t$ (red points).
d) $w$ as a function of $Q_p$ at $t=-1.48$ (blue points) and $t=-1.30\,\sigma_t$ (red points).
The error bars indicate the standard deviation of $w$, and of $Q_e$ and $Q_p$.
Note: blue line: same data on (a) and (c). %
}
\label{fig:4}
\end{figure}

 Figure~\ref{fig:4}(a) shows that, when increasing the charge of the seed bunch $Q_e$, the width $w$ along the bunch initially decreases, following the same curve in each case, due to the effect of adiabatic focusing.
It then increases with SM growth, reaching the value of the case without plasma ($w_{\mathrm{off}}$) earlier along the bunch, for larger $Q_e$ (red points).
Since the global focusing effect is equal in all cases (${Q_p=const}$), this shows that an increase in $Q_e$ causes the SM defocusing effect to dominate earlier along the bunch.
Afterwards, $w$ increases monotonically and reaches larger values at all times for larger $Q_e$, as shown in Fig.~\ref{fig:4}(b) for two times along the bunch (blue points: $t=-1.19\,\sigma_t$, red points: $t=-0.84\,\sigma_t$; first and last $t$ when $w>w_{\mathrm{off}}$ for $Q_e>0$ and all measurements provide a value).
\par Measurement of the energy spectrum of the seed $e^-$ bunch (not shown)~\cite{LIVIO:EPS}, and numerical simulation results~\cite{KJ:PREP} indicate that the amplitude of the wakefields driven by the $e^-$ bunch W$_{\perp 0}$ over the first $\sim 2\,$m of plasma increases as a function of $Q_e$ and exceeds $4\,$MV/m in all cases. %
The earlier occurrence of SM defocusing and the increase in $w$ at all times when $w\geq w_{\mathrm{off}}$ for larger $Q_e$ are thus directly caused by the increase in amplitude of the seed wakefields $W_{\perp0}(Q_e)$, since all other parameters were kept constant ($\Gamma(Q_p)=const$). %
Figure~\ref{fig:4}(a) also shows that in the SMI regime ($Q_e=0$) the defocusing effect of SM dominates much later along the bunch ($\sim -0.78\,\sigma_t$) and $w$ is much smaller than in the seeded regime ($Q_e>0$). 
This lower growth can be attributed to the lower amplitude of the (uncontrolled) initial wakefields, as well as to a later start of SM along the bunch~\cite{FABIAN:PRL}. 

\par When increasing the $p^+$ bunch charge $Q_p$ (Fig.~\ref{fig:4}(c)), we observe again that $w$ increases at all times along the bunch when SM defocusing effect dominates, as also shown in Fig.~\ref{fig:4}(d) for two times along the bunch (blue points: $t=-1.48\,\sigma_t$, red points: $t=-1.30\,\sigma_t$; $t$ chosen as in the previous case).
Increasing $Q_p$ also increases the emittance, transverse size and bunch density $n_p$ of the $p^+$ bunch at the plasma entrance~\cite{pbparams}, and also $w_{\mathrm{off}}$ at the screen (red points, Fig.(c)). %
Therefore, and unlike with $Q_e$, when increasing $Q_p$ the effect of adiabatic focusing ($\propto{n_p}$)
also increases. %
However, measurements show that the increase in $w$ with $Q_p$ is even larger and thus the effect of SM defocusing starts dominating earlier along the bunch. %

\par The expected variation of $\Gamma$ with $n_p$ is $\Gamma\propto n_p^{1/3}$~\cite{KUMAR:GROWTH,PUKHOV:GROWTH,SCH:GROWTH}.
Measurements of $\sigma_{x,y}$ and $\sigma_t$~\cite{pbparams} show that over the $Q_p=(14.7-46.9)\,$nC range, $n_p$ changes only from $6.9$ to $8.9\cdot 10^{12}\,$cm$^{-3}$.
However, the effect of this change is observed after exponentiation of SM.

\par The effect of the increase in transverse size and emittance of the $p^+$ bunch when increasing $Q_p$~\cite{pbparams} (not explicitly included in $\Gamma$) is to reduce the growth of SM~\cite{GORN:DEFOCUSING,LOTOV:EMITTANCE}.
Thus, the increase in $\Gamma$ with $Q_p$ is likely larger than the increase in $w$ shown by Fig.~\ref{fig:4}(d). %

\par We note here that the measurement of $w$ is not direct measurement of the amplitude of the seed wakefields $W_{\perp 0}$ or growth rate $\Gamma$. %
However, changes in $w$ are direct consequences of changes in $W_{\perp 0}(Q_e)$ and $\Gamma(Q_p)$. %
For a direct measurement of $\Gamma$ all protons would have to leave the wakefields at the same position along the plasma and propagate ballistically an equal distance to the OTR screen. %
Numerical simulation results show that with the plasma of these experiments longer than the saturation length of SM~\cite{MARLENE:PRAB}, protons may leave the wakefields earlier or later depending on the amplitude of the wakefields and on the distance they are subject to them. %
However, simulations also show monotonic increase of $w$, as observed in the experiments, and that $w$ increases with increasing amplitude of the wakefields along the bunch. %

\par \textit{Summary.---} We demonstrated in experiments that a short $e^-$ bunch can seed SM of a long $p^+$ bunch in plasma.
We showed that when increasing the $e^-$ ($Q_e$) or the $p^+$ ($Q_p$) bunch charge, the transverse extent of the $p^+$ bunch distribution $w$ along the bunch (measured after the plasma) also increases.
We attribute these changes to the change in amplitude of the seed wakefields ($Q_e\rightarrow W_{\perp 0}$) and in growth rate of SM ($Q_p\rightarrow\Gamma$), in agreement with theoretical and simulation results. %

\par These results show that SM is well understood and can be well controlled. %
Control is key for optimization of the SM wakefields for particle acceleration~\cite{veronica,PATRIC:EAAC}. %

\begin{acknowledgments}
This work was supported in parts by a Leverhulme Trust Research Project Grant RPG-2017-143 and by STFC (AWAKE-UK, Cockcroft Institute core, John Adams Institute core, and UCL consolidated grants), United Kingdom;
the National Research Foundation of Korea (Nos.\ NRF-2016R1A5A1013277 and NRF-2020R1A2C1010835);
the Wolfgang Gentner Programme of the German Federal Ministry of Education and Research (grant no.\ 05E15CHA);
M. Wing acknowledges the support of DESY, Hamburg.
Support of the National Office for Research, Development and Innovation (NKFIH) under contract numbers 2019-2.1.6-NEMZ\_KI-2019-00004 and MEC\_R-140947, and the support of the Wigner Datacenter Cloud facility through the Awakelaser project is acknowledged.
The work of V. Hafych has been supported by the European Union's Framework Programme for Research and Innovation Horizon 2020 (2014--2020) under the Marie Sklodowska-Curie Grant Agreement No.\ 765710.
TRIUMF contribution is supported by NSERC of Canada.
The AWAKE collaboration acknowledge the SPS team for their excellent proton delivery.
\end{acknowledgments}

\bibliography{apssamp}
\end{document}